\documentclass[onecolumn]{emulateapj}
\usepackage{graphics}
\usepackage{color}
\usepackage{hyperref}

\usepackage[colorinlistoftodos]{todonotes}
\newcommand{\smallfoot}[1]{\let\thefootnote\relax\footnotetext{\scriptsize{#1}}}

\definecolor{darkgreen}{rgb}{0,0.5,0}
\definecolor{darkblue}{rgb}{0,0,0.75}
\definecolor{royalblue}{RGB}{0,0,128}
\definecolor{darkred}{RGB}{210,0,0}

\newcommand{\um}{$\rm\mu m$}
\newcommand{\spherex}{SPHEREx}
\newcommand{\panstarrs}{\mbox{Pan-STARRS}}
\newcommand{\fitcat}{\texttt{fitcat}}
\newcommand{\makegrid}{\texttt{\mbox{make-grid}}}
\newcommand{\photoz}{\texttt{\mbox{photo-z}}}
\newcommand{\fnl}{$f_{\rm NL}$}

\shorttitle{An Empirical Approach to Cosmological Galaxy Survey Simulation}
\shortauthors{N.~R. Stickley et al.}

\begin{abstract}
 Highly accurate models of the galaxy population over cosmological volumes are
necessary in order to predict the performance of upcoming cosmological missions. We
present a data-driven model of the galaxy population constrained by deep 0.1--8~\um\
imaging and spectroscopic data in the COSMOS survey, with the immediate goal of simulating
the spectroscopic redshift performance of the proposed SPHEREx
mission. SPHEREx will obtain over the full-sky $R\sim41$
spectrophotometry at moderate spatial resolution ($\sim6$\arcsec) over
the wavelength range 0.75--4.18~\um\ and $R\sim135$ over the
wavelength range 4.18--5~\um\ .  We show that our simulation accurately
reproduces a range of known galaxy properties, encapsulating the full complexity of the
galaxy population and enables realistic, full end-to-end simulations to predict
mission performance.  Finally, we discuss potential applications of
the simulation framework to future cosmology missions and give a
description of released data products.
\end{abstract}


\begin{document}

\title{An Empirical Approach to Cosmological Galaxy Survey Simulation: Application to
SPHERE\MakeLowercase{x} Low Resolution Spectroscopy}

\author{Nathaniel R. Stickley\altaffilmark{1,3}, Peter
  Capak\altaffilmark{2}, Daniel~Masters \altaffilmark{2},
  Roland~de~Putter\altaffilmark{3},
  Olivier~Dor\'{e}\altaffilmark{4,3}, Jamie~Bock \altaffilmark{3}}

\altaffiltext{1}{Department of Physics and Astronomy, University of  California,Riverside, USA}
\altaffiltext{2}{Spitzer Science Center, California Institute of Technology, Pasadena, CA 91125, USA}
\altaffiltext{3}{California Institute of Technology, Mail Code 350-17, Pasadena, California 91125, USA}
\altaffiltext{4}{Jet Propulsion Laboratory, California Institute of Technology, Pasadena, California
91109, USA}

\section{Introduction} \label{section:introduction}

Low-resolution ($R\sim20$--100) spectroscopy and photometric redshifts (photo-z's) based on
$R\lesssim10$ broad-band photometry are now widely accepted as powerful tools for galaxy evolution
studies and precision cosmology. Current surveys such as CANDELS \citep{candels}, COSMOS
\citep{scoville2007}, ALHAMBRA \citep{alhambra}, and PRIMUS \citep{primus1, primus2} have
successfully used $R\sim5$--100 spectrophotometry to estimate redshifts for a wide variety of
sources with great success. Furthermore, upcoming ``Stage~IV'' dark energy experiments such as
LSST,
\emph{Euclid}, and \emph{WFIRST} will rely critically on photometric redshifts for weak lensing
cosmology.

A key shortcoming in galaxy simulations aimed at informing cosmology missions has been that the
photo-z performance predicted by such simulations is typically overestimated compared to actual
performance.  This is largely due the inherent complexity of the galaxy population---which is
rarely
captured in simulations---combined with the myriad systematic effects that increase the noise in
spectrophotometry and are difficult to model without full end-to-end simulations. Cosmological
simulations based on the actual, measured galaxy population that account for all sources of
measurement errors would therefore represent a major improvement in terms of capturing the true
complexity that cosmological surveys would encounter.

In this paper, we present a simulation framework aimed at capturing the full complexity of the
galaxy population using deep, 30-band data in the well-studied COSMOS field
\citep{scoville2007,laigle2016}.  This field includes very deep 0.1--8~\um\ data with spectral
resolution of $R\sim5$--20 as well as X-ray and longer wavelength data. In addition, high-quality
redshifts are available to $\sim$25,000 galaxies, and a new survey is underway to obtain spectra
fully representative of the galaxy population (Masters et al. in prep).  Furthermore,
photometric redshifts accurate to $dz/(1+z) < 0.01$ are available for most galaxies. With these
data, the spectral energy distributions (SEDs) of the galaxies are very well constrained, so the
fluxes in unknown bands in the 0.1--8~\um\ range can be estimated to high-precision via
\emph{interpolation}, rather than extrapolation.  The resulting simulation accurately reproduces
narrow and intermediate band photometry not used to generate the model. A wide range of galaxy
population properties, such as the stellar masses, star formation rates, and emission line
characteristics of galaxies are thereby accurately reproduced.

In addition to the SEDs, high-resolution \emph{Hubble Space Telescope} (HST) imaging and galaxy
physical property (star formation rate, stellar mass) measurements are available in the COSMOS
field
\citep{laigle2016}.  The galaxies are also tied to their dark matter halo properties using a
combination of weak lensing, clustering, and abundance matching (Leauthaud et al. 2012). With this
information, it is straightforward to propagate the interpolative photometry model to cosmological
constraints.

We apply this simulation framework to an end-to-end simulation of the proposed \spherex\
mission\footnote{\url{http://spherex.caltech.edu}}, which will obtain $R\sim40$ spectrophotometry
over the wavelength range 0.75--4.18~\um\ and $R\sim135$ over the wavelength range 4.18--5~\um\
for
the full sky at moderate (6.2\arcsec) spatial resolution \citep{Dore:2014}. The simulation includes
source confusion, selection effects, systematic calibration errors, and photometric extraction
errors. Using this end-to-end simulation, we realistically constrain the expected redshift
performance of the proposed \spherex\ mission, and the resulting cosmological constraints.

The outline of this paper is as follows. In \S\ref{section:philosophy}, we outline the simulation
philosophy. In \S\ref{section:spherex-overview}, we give an overview of the proposed \spherex\
mission. In \S\ref{section:simulating-spherex}, we present the methodology used to generate
realistic mock \spherex\ data based on the COSMOS survey. In \S\ref{section:pipeline-overview} we
describe the simulation pipeline in detail.
In \S\ref{section:conclusions}, we conclude with a discussion of how similar simulations may
benefit
a number of upcoming cosmology missions, such as LSST \citep{SciBook}, \emph{Euclid}
\citep{Laureijs:2011gra}, and \emph{WFIRST} \citep{Spergel:2015sza}. We also discuss potential
applications to predicting the performance of spectroscopic surveys such as PFS \cite{Takada:2014}
and DESI \citep{Levi:2013}.

\section{Simulation Philosophy} \label{section:philosophy}

The key idea behind the \spherex\ simulation is to stay as close to reality as possible
by using the well-constrained galaxy photometry in the COSMOS field to generate realistic
simulated photometry. Rather than try to simulate the dark matter structure of the
universe and paint on galaxies, we let the observed universe in the COSMOS field drive the
simulated observations. In this way, we ensure that the simulated data captures the
complexity of the real world.

Galaxies in the COSMOS field have deep, 30-band photometric data and accurate photometric redshifts,
in addition to other measured quantities. To predict the \spherex\ observations, intrinsic spectral
energy distributions (SEDs) of the galaxies must be assumed. We use the state-of-the-art galaxy
spectral templates from \citet{browntemplates}, an empirically-based set of SEDs representing a wide
range of galaxy types, to assign a realistic SED to each galaxy in COSMOS, including emission lines.
This is accomplised by fitting the templates to the observed COSMOS photometry, which is highly
constraining over the relevant wavelength range. This data-driven SED then feeds forward into the
simulated \spherex\ photometry. Critically, the effects of blended photometry and the realistic
distribution of galaxy properties in the universe are automatically modeled with this approach.

\section{\spherex\ Overview} \label{section:spherex-overview}

\spherex\ (Spectro-Photometer for the History of the Universe, Epoch of Reionization, and Ices
Explorer) is a proposed all-sky survey satellite designed to address all three science
goals in NASA's Astrophysics Division: probe the origin and destiny of our Universe; explore whether
planets around other stars could harbor life; and explore the origin and evolution of galaxies.
\spherex\ will probe the origin of the Universe by constraining the physics of inflation, the
superluminal expansion of the Universe that took place some
$10^{-32}$~s after the Big Bang \citep{Dore:2014}. \spherex\ will
study the imprint of inflation in the three-dimensional large-scale
distribution of matter by measuring galaxy redshifts over a large cosmological volume at low
redshifts, complementing
high-redshift surveys optimized to constrain dark energy. \spherex\ will also investigate the
origin of water and biogenic molecules in all phases of planetary system formation---from molecular
clouds to young stellar systems with protoplanetary disks---by measuring absorption spectra to
determine the abundance and composition of ices toward $> 2\times
10^4$ Galactic targets. \spherex\ will chart the origin and history of
galaxy formation through a deep survey mapping large-scale
structure. This technique measures the total light produced by all
galaxy populations, complementing studies based on deep galaxy counts,
to trace the history of galactic light production from the present day
to the first galaxies that ended the cosmic dark ages. \spherex\ will
be the first all-sky near-infrared spectral survey, creating a legacy archive of spectra ($0.75 \leq
\lambda \leq 4.18$~\um\ with $\lambda/\Delta\lambda = 41.4$, and a narrower filter width
$\lambda/\Delta\lambda = 135$ in the range $\lambda = 4.18$--$5$~\um) with the high sensitivity
obtained using a cooled telescope 4with large spectral mapping speed.

The \spherex\ Mission will implement a simple, robust design that maximizes spectral throughput and
efficiency. The instrument is based on a 20~cm all-aluminum telescope with a wide
$3.5^\circ\times7^\circ$ field of view, imaged onto four $\rm 2k \times 2k$~HgCdTe detector arrays
arranged in pairs, separated by a dichroic. Spectra are produced by four linear-variable filters
(LVFs). The spectrum of each source is obtained by moving the telescope across the dispersion
direction of the filter in a series of discrete steps---a method demonstrated by LEISA on New
Horizons \citep{newhorizons}. \spherex\ has no moving parts except for one-time deployments of the
sun shields and aperture cover. It will observe from a sun-synchronous terminator low-earth orbit,
scanning repeatedly to cover the entire sky in a manner similar to IRAS \citep{iras}, COBE
\citep{cobe} and WISE \citep{wise}. During its two-year nominal mission, \spherex\ produces four
complete all-sky spectral maps for constraining the physics of inflation. The orbit naturally covers
two deep, highly redundant regions at the celestial poles, which we use to make a deep map, ideal
for studying galaxy evolution and monitor instrument performances.

\section{Simulating the \spherex\ Data using the COSMOS Survey} \label{section:simulating-spherex}

To accurately assess the ability of \spherex\ to recover redshifts, the simulation must include the
full observed diversity in both the broad SEDs and emission line features of galaxies. Furthermore,
blends of objects and artifacts that produce spectra with biased redshifts must also be simulated
to
accurately represent the redshift ambiguity in the data.

To ensure this level of representativeness, we use data from the Cosmic Evolution Survey (COSMOS,
\citealp{scoville2007, capak2007}). COSMOS is a multi-waveband survey of a 2-deg$^{2}$ patch
of equatorial sky with observations from the X-ray to the radio. COSMOS is built around deep
\emph{HST}-ACS imaging in $i$-band, with imaging in a number of ground-based telescopes and other
major space-based observatories providing broad and intermediate band imaging over 0.1--8~\um\ at a
resolution of $R=5$--20.  COSMOS therefore covers the \spherex\ wavelength range at similar (though
slightly lower) spectral resolution, but is more than 100 times as sensitive and has more than five
times the spatial resolution. Furthermore, a fully representative spectroscopic data set is
available at the \spherex\ flux limits in COSMOS.

To interpolate to the higher spectral resolution of \spherex, we use the observed Galaxy templates
from \citet{browntemplates}, AGN \citep{salvato2011}, and stars \citet{Chabrier:2000}, which span
the range of expected object properties and include all of the key features we will observe with
\spherex. We note that this pipeline is an extension of the pipeline used to produce official
forecasts for the \emph{Euclid} consortium, the \emph{WFIRST} Science Definition Team, as well as
the Hyper Suprime-Cam (HSC) Subaru survey.

\section{Simulation Pipeline Overview} \label{section:pipeline-overview}

There are four main steps to the \spherex\ simulation pipeline: 1) Create an input galaxy
catalog, based on real data from the multi-wavelength COSMOS survey. 2) Generate simulated
\spherex\ images at each wavelength step of the \spherex\ LVFs. 3) Optimally extract the photometry
from the simulated \spherex\ images. 4) Derive the photometric redshifts for all detected objects.
In this section, we describe these steps in detail.


\subsection{Galaxy redshifts and spectral energy distributions from COSMOS}
\label{section:pipeline-input-data}

We use the COSMOS 2-deg$^2$ \citet{laigle2016} COSMOS multi-band catalog as the starting point of
our simulation. The catalog includes some of the deepest available spectro-photometry from 0.1~\um\
to 8~\um. It was selected at 0.9--2.5~\um\ (similar to \spherex), using a combination of Subaru
Suprime-Cam and UltraVISTA DR2 imaging \citep{mccracken2013}. The catalog is 95\% complete to ${\rm
z}'<25.5$ (0.9~\um) and ${\rm Ks}<24.5$ (2.5~\um) AB magnitudes and includes data of equivalent
depth from 0.1–8~\um. The \citet{laigle2016} data have been shown to produce photometric redshifts
precise to $\sigma(z)/(1+z)<0.02$ with a very low outlier fraction at \spherex\ depths. Furthermore,
the
accuracy, precision, and other properties of the photometric redshifts has been verified with a
representative spectroscopic sample at \spherex\ depths \citep{masters2015}. As a result,
for all objects in the COSMOS field detectable by \spherex, the observed SEDs and redshifts are
known. The observational data in COSMOS reach two orders of magnitude fainter in flux
than will be detectable by \spherex, so the effects of confusion and blending can be accurately
modeled in our simulations.

The spectroscopic catalog produced from this photometry includes a classifier for galaxies, AGN, and
stars based on the full photometric data set along with X-ray and radio data to select AGN. We
acknowledge the fact that there are more robust classifiers for AGN and better redshift
determinations
than \citet{laigle2016} \citep[for instance,][]{salvato2011}. However, for the purposes of
simulating the photometry and the performance of spectroscopic
redshift or photo-z, the exact classification and redshift
distribution has little effect. Specifically, the main impact of AGN on photo-z performance is the
template confusion and the method outlined in this paper accurately reproduces that effect.

\begin{figure}

\centering

\includegraphics[width=0.45\textwidth, natwidth=7.09in,
natheight=5.59in]{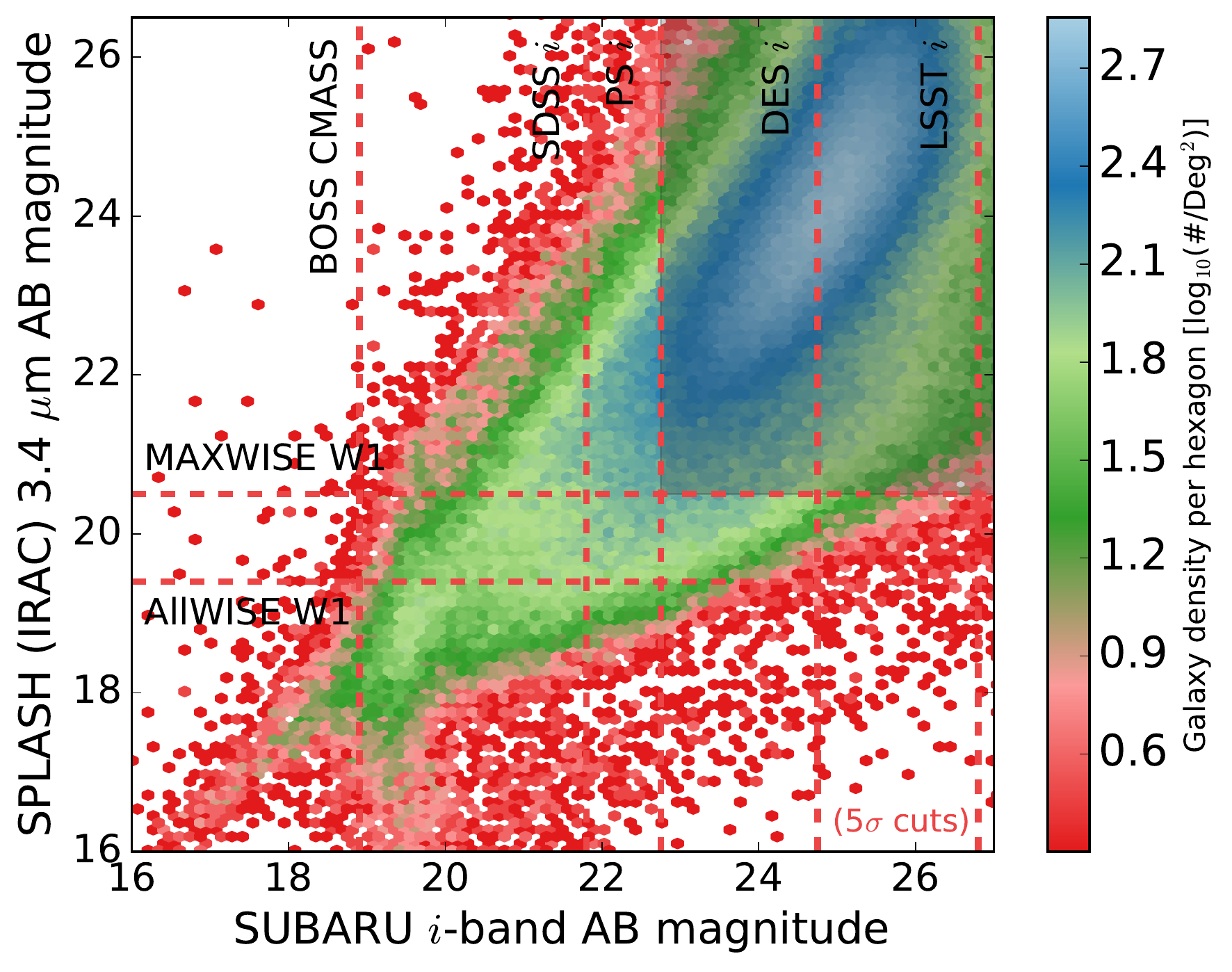}
\includegraphics[width=0.45\textwidth, natwidth=7.10in,
natheight=5.71in]{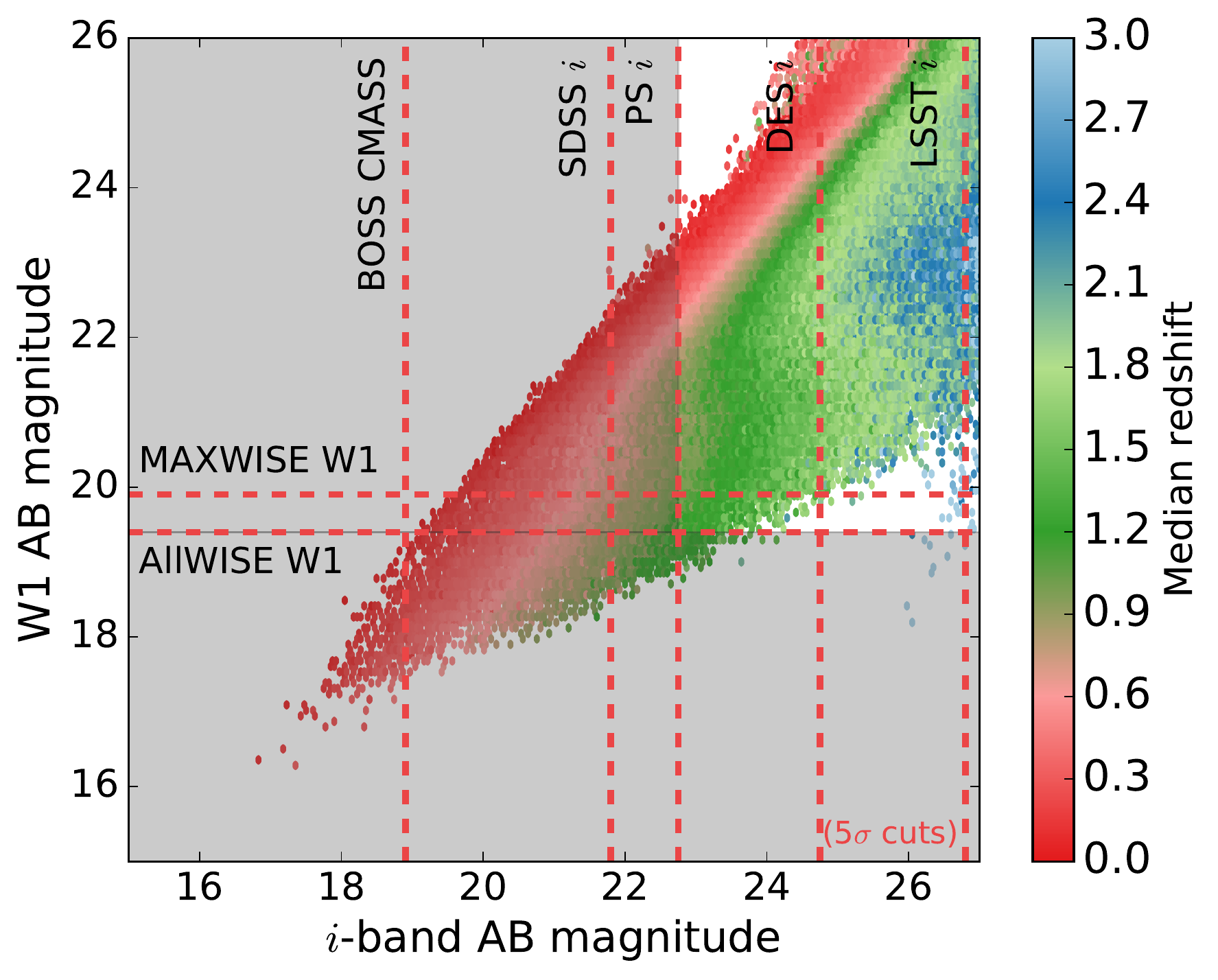}

\caption{\label{figure:phot-comparison} Magnitude-magnitude plots of
  the COSMOS data. In both panels, the vertical and horizontal
lines indicate 5 $\sigma$ depths for various relevant surveys /
instruments . All the objects in this plot are included in our
simulations. The gray shaded region identify objects that contribute
to the confusion background in the simulations but are not selected to be in
the \spherex\ catalog as being not bright enough in WISE or
PanSTARRS. \textbf{Left panel}: The actual COSMOS catalog in the IRAC
channel~1 $i$-band color space from the Spitzer survey and the Subaru telescope (by
design, IRAC channel~1 is a good proxy for the WISE W1
band). Photometric artefacts appearing around $i$-band M$_{AB}\sim$ 19
are not corrected. The color of the hexagonal bins encodes the
  angular density of detected objects. \textbf{Right panel}: Simulated data from the based
simulation in the WISE W1--$i$ color space. The color of the hexagonal bins encodes the
  median redshift within a bin.}

\end{figure}

Figure~\ref{figure:phot-comparison} demonstrates how well the simulation reproduces the scatter in
the COSMOS data. The \spherex\ spectroscopic catalog is represented by
the non gray shaded area (i.e., the objects detected in \panstarrs\
PS1 $i$-band or WISE~W1). It also demonstrates the fact that
the simulation is based on a data set that is much deeper than the \spherex\ depth, thus allowing
an accurate modeling of confusion and blending.

\subsection{Determination of \spherex\ fluxes} \label{section:pipeline-input-fluxes}

\begin{figure}

\centering

\includegraphics[width=6in,natwidth=11in, natheight=8.5in]{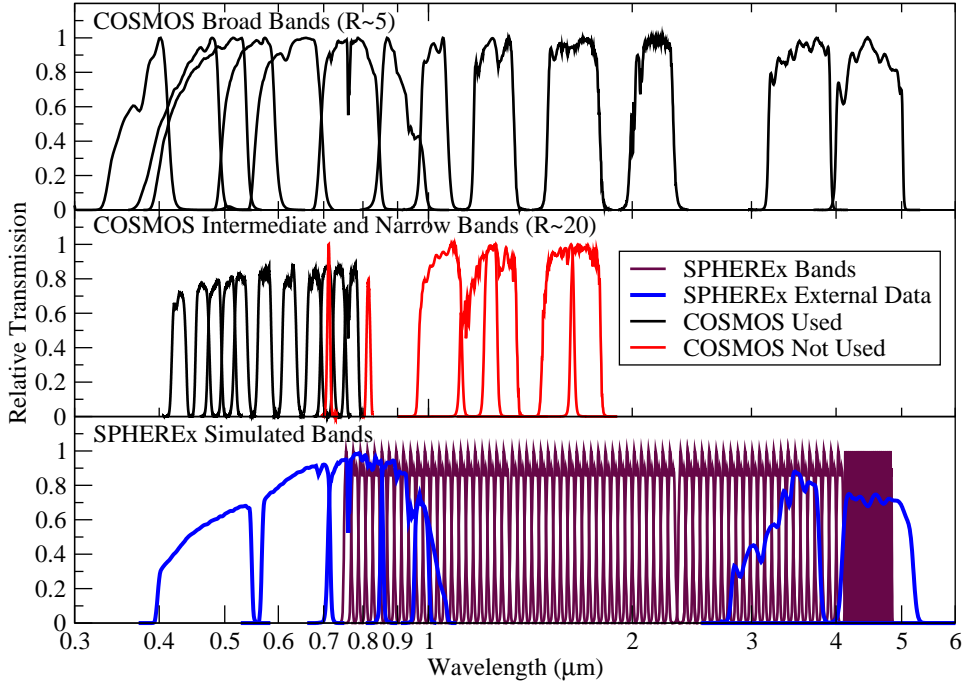}

\caption{\label{figure:filters}The COSMOS filters used to constrain the Brown et al. templates are
shown in black (COSMOS Used), however the GALEX filters at 0.15$\mu$m and 0.25$\mu$m have been
omitted for clarity. COSMOS filters not used to constrain the templates whose fluxes were
predicted as a verification of the method are shown in red (COSMOS Not Used). The \spherex\ bands
on the spacecraft are shown in brown (\spherex\ Bands), while the external data from PS1 and WISE
are shown in blue (\spherex\ external).}

\end{figure}

\spherex\ will reach $R\sim41.5$ with continuous coverage from 0.7--4.5~\um. However, the COSMOS
photometry in the 0.7--5~\um\  range is $R\sim5$--20 and has gaps in the 1--3~\um\  range, due to
atmospheric absorption (See Figure~\ref{figure:filters}), so we must interpolate from the COSMOS
data to \spherex\ spectral resolution. Despite the fact that some spectral information is missing,
SED templates are highly constrained by the COSMOS data, due to its depth, spectral resolution, and
spectral coverage. The performance of the simulation is not highly sensitive to the choice of
template set because the interpolation between bands is relatively small. The most important aspect
of the choice of template library is the diversity of high-resolution ($R>10$) spectral properties
between 1--3~\um, which are not completely constrained by the $R\sim5$ photometry. For galaxies, we
chose the \citet{browntemplates} observed 0.1--100~\um\  templates for the interpolation. These
templates were selected to represent the range of colors observed in local galaxies, and are
constructed from observed spectra whenever possible and photometry when spectra were unavailable.
They include the effects of extinction, complex morphology, and a range of emission line strengths.
We augment the library by adding additional extinction to the templates to represent
heavily obscured sources. The \citet{calzetti2000}, \citet{prevot1984}, \citet{fitzpatrick1986},
\citet{seaton1979}, and \citet{allen1976} laws were used, with additional extinction of up to an
$E(B-V)$ of 1.0 added. In addition to the \citet{browntemplates} templates, the \citet{salvato2011}
templates were used to represent known AGN in the COSMOS field. These templates were chosen to
represent the range of known AGN from broad line QSO to liners and, like the \citet{browntemplates}
templates, are tied to observed spectra and photometry. Finally, for
stars the \citet{Chabrier:2000} library was used.

We note that several other template sets were considered for the interpolation. These included the
\citet{bruzual-charlot} templates, the \citet{ilbert2009} templates, and the MAGPHYS \citep{magphys}
templates. The \citet{bruzual-charlot} templates were rejected because they included neither dust
emission, which dominates the SEDs at $>1$~\um, nor emission lines, which will be a significant
effect at the spectral resolution of \spherex. The \citet{ilbert2009} templates were used in early
iterations of these simulations, but were replaced because they were not fully representative of the
diversity of emission lines and $>1$~\um\  SEDs observed in galaxies. Finally, the MAGPHYS templates
were considered, but were found to be inferior to the \citet{browntemplates} templates because they
did not include optical or NIR emission lines and were not as well-constrained by observations.

In order to fit templates to the COSMOS photometry, we first partitioned the catalog, based
on the \citet{laigle2016} classifications, to create separate galaxy, AGN, and stellar catalogs.
For each galaxy and AGN SED, we modified the templates in the respective template libraries (Brown
et al. templates for galaxies and Savato templates for AGNs) so that the redshifts of the templates
matched the estimated redshifts from \citet{laigle2016}. We modified the extinction of each
template, as discussed above, and then identified the combination of template, reddening law, and
$E(B-V)$ which best fit the photometry (i.e., the combination of parameters which minimized the
$\chi^2$ discrepancy between the model SED and the measured photometry). In the remainder of this
paper, we refer to the combination of template, redshift, reddening law, and $E(B-V)$ as a
\textit{model SED}. For stars, we simply assigned the best-fitting \citet{Chabrier:2000} template.

The best-fitting model SED was then used to interpolate from the COSMOS photometry to the \spherex\
bands using the LVF filter profiles, spaced at $R=41.5$ with Nyquist sampling. The \panstarrs\ PS1
\citep{panstarrs2002, panstarrs2004} and ALLWISE band fluxes are also estimated.
COSMOS


\subsection{\spherex\ image simulation} \label{section:pipeline-images}

\begin{figure}

\centering

\includegraphics[width=4in,natwidth=3.84in, natheight=3.84in]{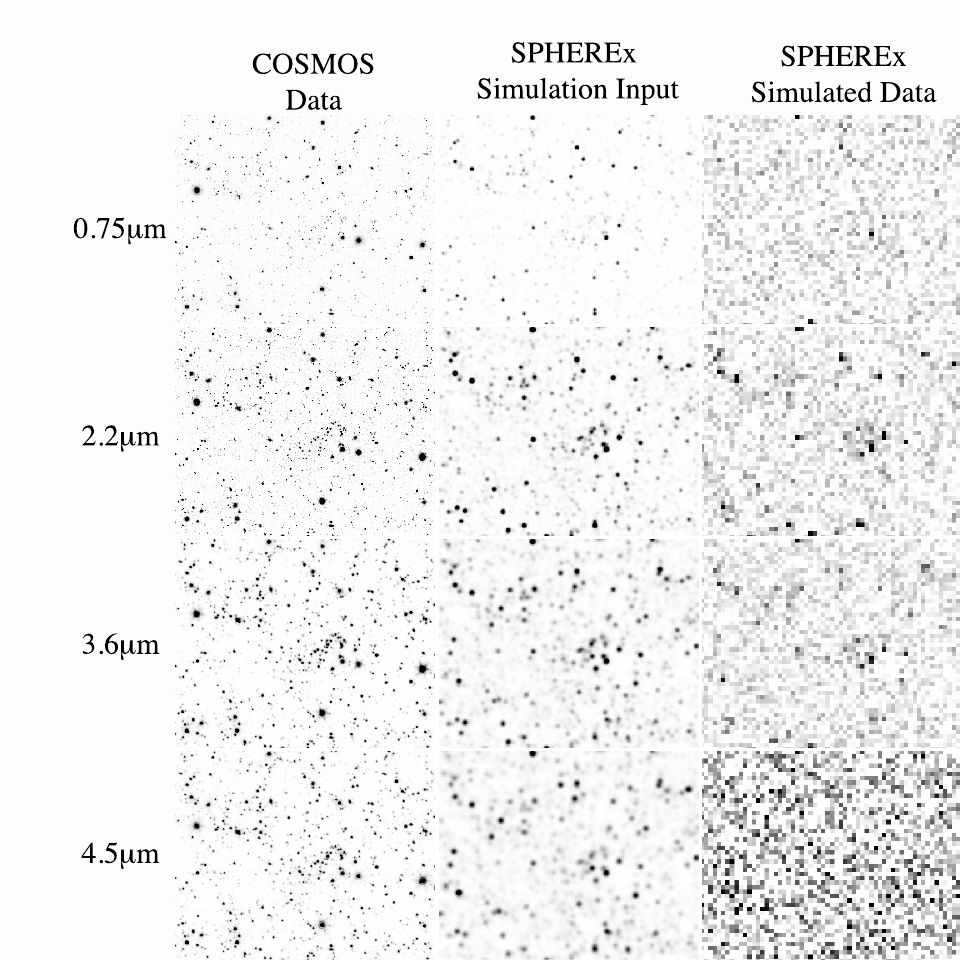}

\caption{\label{figure:images} A comparison of real and simulated images in four bands. Each
sub-plot shows the $5'\times5'$ region at the center of the COSMOS field. The left panels
show real observational data. The center panels show the high-resolution, noiseless simulated
images generated by the pipeline, and the right panels show the lower resolution, final images
after noise has been added.}

\end{figure}

To simulate the effects of confusion, calibration, and extraction on the photometry we constructed
an end-to-end simulation of the images, photometric extraction, and redshift recovery. We begin with
the predicted fluxes in \S\ref{section:pipeline-input-fluxes}. These fluxes are then added to a set
of simulated images (one image for each \spherex\ band), oversampled by a factor of 8 relative to
the \spherex\ detector (i.e., to a grid with pixel scale 6$\farcs$2/8). All sources in the
\citet{laigle2016} catalog are placed in the simulated images, reaching to over 100 times the
detection limit of \spherex.

These images are then smoothed with a Gaussian of $1.22 \lambda/D$ to simulate diffraction, a
1.55\arcsec\ FWHM Gaussian (4.5~\um\  scaled to the plate scale of the detector) to simulate optical
aberration, which dominates over diffraction (by design) for much of the wavelength coverage. A
8.4''\arcsec\ FWHM Gaussian to simulate pointing error. The image is then binned to the \spherex\
pixel scale. Based on instrument simulations, a random multiplicative error of 0.32\% is applied to
each pixel to represent the worst case flat fielding error and an additive error of between
2--4~$\rm\mu Jy$ (depending on wavelength) to represent sky background subtraction error. Finally,
Gaussian noise is added at the expected RMS background level.

\subsection{Source extraction and sample properties} \label{section:pipeline-photometry}

To extract photometry, we use the \panstarrs\~PS1 and ALLWISE catalogs for source identification and
astrometry. A median background is subtracted from all sources to correct for the mean contribution
of undetected/unknown sources. The \spherex\ PSF at the object position is used to perform optimal
flux extraction, taking into account the pixel response function of both the primary pixel on which
the object falls as well as neighboring pixels.  This is done for each \spherex\ band.

For an unresolved source in a survey with a well-known PSF, $P$, the optimal photometric extraction
of the total flux, $F$, is obtained by the weighted sum,
\begin{equation}
F = \sum_{ij}w_{ij}D_{ij},
\end{equation}
where $D_{ij}$ is the sky-subtracted flux measured in pixel $(i,j)$ and the weight function $w_{ij}$
is
\begin{equation}
w_{ij} = \frac{P_{ij}}{\sum_{ij}P^{2}_{ij}}.
\end{equation}
Here $P_{ij}$ is the fraction of the PSF falling in pixel $(i,j)$. In the all-sky survey proposed
for \spherex\, the noise budget is dominated by photon noise from zodiacal light, which is diffuse
and nearly uniform across the FOV.  To quantify the signal-to-noise of a given detection and how it
varies with the relative alignment of the detector grid, it is useful to define the parameter
\begin{equation}
N_{\rm eff} = \sum_{ij}P_{ij}^{-2},
\end{equation}
which represents the effective number of pixels spanned by the PSF. As $N_{\rm eff}$ increases, the
signal from the source is spread over more pixels and the noise increases by the square root of the
number of pixels.  Therefore, in optimal photometry, the uncertainty in recovered source flux is
related to $N_{\rm eff}$ by
\begin{equation}
\delta F=  \delta F(1) \sqrt{N_{\rm eff}}
\end{equation}
where $\delta F(1)$ is the uncertainty in flux achieved for a PSF which is a perfect single-pixel
square step function.

For \spherex, $N_{\rm eff}$ will increase systematically with wavelength, due to the growth of
the diffraction limited PSF. Additionally, for a single wavelength, there is a significant spread in
$N_{\rm eff}$, caused by the random alignment of sources with the detector grid. If a source falls
closer to the corner of a pixel, its flux will spill over to neighboring pixels, while a source
landing in the center will deposit most of its energy in a single pixel.

\begin{figure}

\centering

\includegraphics[width=3in,natwidth=4.0in, natheight=3.8in]{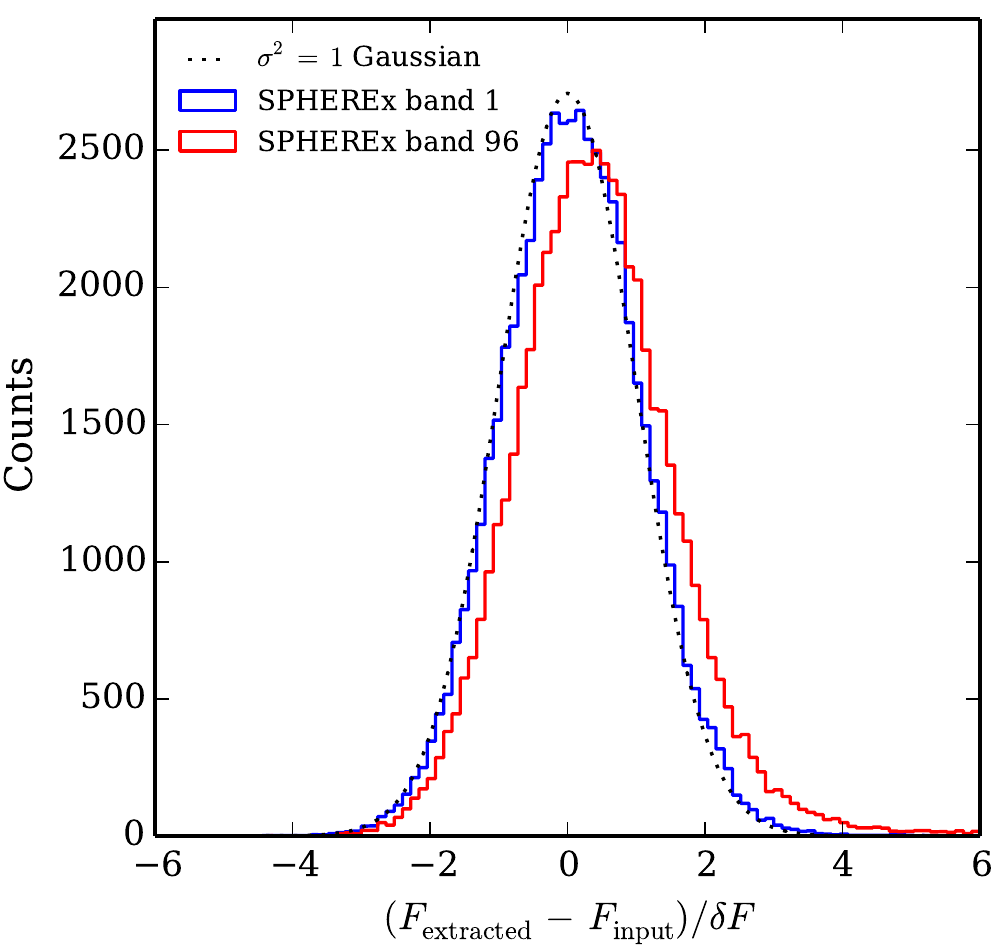}

\caption{\label{figure:flux-error} Histograms showing the relationship between the input fluxes
($F_{\rm input}$) and fluxes extracted from synthetic images ($F_{\rm extracted}$). The dotted
black line shows the ideal case, in which the extracted fluxes are perfectly consistent with the
input fluxes, after taking the measurement errors ($\delta F$) into account. The blue and red
histograms show the shortest and longest wavelength bands of \spherex, respectively. Contamination
from neighboring objects tends to cause a positive offset in the extracted fluxes. Longer
wavelengths are more strongly affected than shorter wavelenths because the PSF increases with
wavelength.}

\end{figure}

With the position of the object known, from \panstarrs\ and ALLWISE, the absolute pointing
reconstructed to $< 1$\arcsec, and the PSF accurately characterized in every image, the values of
$P_{ij}$ that go into the weight function for optimal extraction can be computed. To do this, the
normalized PSF is resampled onto a fine grid centered at the known position of the object on the
detector, and the value of $P_{ij}$ in the nearby pixels is found by summing the portion of $P$
falling within pixel $(i,j)$.

In general, the optimally extracted object fluxes match the input fluxes well---particularly for
brighter objects in the input catalog. However, there can be significant contamination, in
particular for fainter objects with contaminating objects that fall in the same or neighboring
\spherex\ pixels. The significance of contamination increases as the PSF grows. Thus, contamination
is more significant for the longer wavelenth bands, as illustrated in
Figure~\ref{figure:flux-error}.

\subsection{Redshift estimation} \label{section:photoz-algo}

In addition to simulationg the \spherex\ fluxes, we use the model SEDs to generate simulated
\panstarrs\ ground-based photometry and WISE photometry in the 3.4~$\mu$m (W1) and 4.6~$\mu$m (W2)
bands. Appropriate noise is added to these simulated observations, which represent ancillary data
that will be available to the \spherex\ mission.

The \spherex\ flux data are combined with the simulated \panstarrs\ PS1 + WISE photometry to
construct SEDs for each object. These SEDs are then processed by a redshift estimation code, based
on LePHARE \citep{lephare1, lephare2}, but re-implemented in C++. The code computes the redshift
likelihood distribution, $P(z)$,  for each observed galaxy by comparing the \spherex\ + PS1 + WISE
photometry with a large grid of model SEDs. More specifically, $P(z)$ for each observed object is
computed as
\begin{equation}
    P(z) = \sum_i e^{-\chi_i^2(z)/2}, \label{eq:pdf}
\end{equation}
where the quantity $\chi_i^2(z)$ is a measure of the discrepancy between the observed SED and the
$i$th model SED of redshift, $z$. The sum in Eq.~\ref{eq:pdf} is performed over all model SEDs of
redshift, $z$.

Let $F_j$ and $\delta F_j$ represent the observed flux through the $j$th band and the error in the
flux measurement, respectively. We define the associated weight, $w_j = \delta F_j^{-2}$ and we
denote the flux associated with the $j$th band of the $i$th model SED at redshift, $z$, as
$f_{ij}(z)$. In terms of these quantities, $\chi_i^2(z)$ is given by
\begin{equation}
 \chi_i^2(z) = \sum_j w_j \left[F_j - s_if_{ij}(z) \right]^2,
\end{equation}
where $s_i$ is a scale factor, chosen to minimize $\chi_i^2(z)$ (i.e., the value satisfying the
condition \mbox{$\partial\chi^2_i(z)/ \partial s_i = 0$}):
\begin{equation}
   s_i = \frac{\sum_j w_j F_j f_{ij}(z)}{\sum_j w_j f_{ij}^2(z)}.
\end{equation}
The redshift and redshift error are estimated as the mean and standard deviation of the likelihood
distribution, respectively.  Explicitly,
\begin{equation}
    z_{\rm est} = \left\langle z \right\rangle = \frac{\sum_i z_i P(z_i)}{\sum_i P(z_i)}
\end{equation}
and
\begin{equation}
    \sigma_z = \sqrt{\left\langle z^2 \right\rangle - \left\langle z \right\rangle^2},
\end{equation}
where
\begin{equation}
    \left\langle z^2 \right\rangle = \frac{\sum_i z_i^2 P(z_i)}{\sum_i P(z_i)}.
\end{equation}

\begin{figure}

\centering

\includegraphics[width=4.5in,natwidth=6.0in, natheight=7.0in]{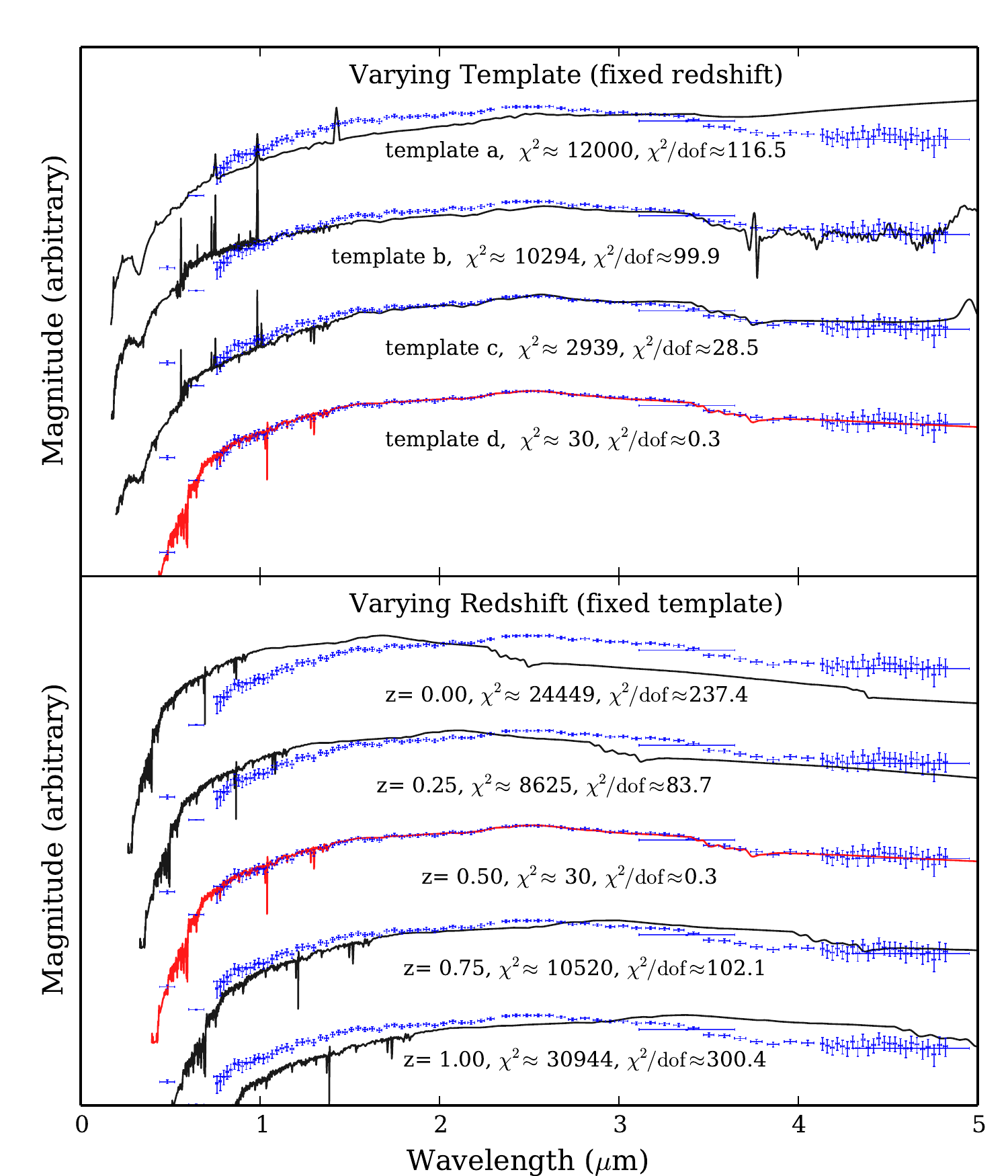}

\caption{\label{figure:fitting-demo}A demonstration of the model-fitting procedure, using a
small subset of the SED models in the grid. The high-resolution version of the model SEDs (thin
lines), are plotted against an observed SED (blue points). The values of $\chi^2$ and $\chi^2$
per degree of freedom ($\chi^2/\rm dof$) are indicated. \textbf{Upper panel}: Four SED models with
constant $z$, $E(B-V)$, and reddening law, but varying base template are compared with the
observed fluxes. The best-fitting model is plotted in red. \textbf{Lower panel}: The template,
$E(B-V)$, and reddening law are held constant, while $z$ is varied. The best-fitting model is
plotted in red.}

\end{figure}

As discussed previously, the model grid is constructed using SED templates from
\citet{browntemplates} and \citet{salvato2011}. For each of the 188 templates in our grid, we sample
redshifts from 0--4.5 in steps of $dz = 0.0015$, $E(B-V)$ values from 0--1 in steps of
$dE(B-V)=0.045$, and five reddening laws, which means the model grid contains 55,146,667 model SEDs.
The expected fluxes of each model in the grid are pre-computed by integrating the high-resolution
model against the \spherex\ LVF filter profiles and the PS1 + WISE filter profiles. These fluxes are
then stored in a file prior to the model-fitting step.

Figure~\ref{figure:fitting-demo} shows a small sample of SED models in the model grid, compared with
an observed SED. We have shown the high-resolution versions of the models so that the spectral
features being probed are more obvious to the eye. The plot demonstrates how drastically $\chi^2$
varies with different model SEDs.

\subsection{Implementation in AWS} \label{section:pipeline-aws}

The simulation code and input data files are stored in an Amazon Machine Image (AMI) so that the
simulation can run in Amazon.com's Elastic Compute Cloud (EC2). The AMI includes the StratOS
software framework \citep{stratos}, which can automatically build a virtual computing cluster from
EC2 instances. The StratOS framework is used to execute the most computationally expensive parts of
the simulation on the slave nodes of the virtual cluster, while less expensive tasks are typically
performed on the master node. For example, in order to efficiently create the grid of model SEDs,
the input data from the COSMOS catalog is partitioned into smaller sections, which are sent to the
slave nodes for processing. While the slave nodes are busy generating the model grid, the master
node creates synthetic images for each band, measures the fluxes from the images, and constructs a
catalog of observed SEDs. For more details on the sub-components of the pipeline, refer to the
Appendix.

The StratOS framework efficiently schedules the tasks running on the slave nodes, which prevents the
simulation from being slowed down significantly by a small number of low-performing nodes (e.g.,
nodes running in EC2 instances that have shared tenancy). Without the sort of scheduling provided by
StratOS, the performance of the simulation would be limited by the speed of the slowest slave node.
The framework's resilience to node failure also prevents work from being lost in the rare event of
an EC2 host failure.



\section{Conclusions} \label{section:conclusions}

Using a detailed end-to-end simulation based on empirical data from the COSMOS survey, we have
verified the ability of the all-sky \spherex\ mission to recover accurate redshifts to a sufficient
number of galaxies to put important new constraints on cosmological parameters, in particular \fnl.
The realism of the simulation, which captures systematic instrument effects as well as confusion and
noise due to galaxies below the \spherex\ detection limit, gives us significant confidence in the
cosmological performance of the proposed \spherex\ mission.

Generally, we have demonstrated that deep, multiwaveband photometry of surveys such as COSMOS can be
used to place highly realistic constraints on the expected performance of cosmology surveys. The
empirically-driven, interpolative approach to simulating mission performance is preferable to pure
simulations in that it captures all of the relevant factors with high fidelity. Important effects
captured by the simulation include source confusion, noise due to faint sources below the detection
limit, realistic galaxy SEDs, and representative emission line characteristics. The framework
developed here for simulating mission performance of wide-area cosmology surveys is broadly
applicable to studies of the performance of future missions such as LSST, \emph{Euclid}, and
\emph{WFIRST}, and can be applied to the problem of predicting the performance of BAO and weak
lensing cosmology with these missions with minor modifications.

\acknowledgments

\section{Acknowledgements} \label{section:acknowledgements}

Part of the research described in this paper was carried out at the Jet Propulsion Laboratory,
California Institute of Technology, under a contract with the National Aeronautics and Space
Administration. RdP and OD acknowledge the generous support of the Heising-Simons Foundation.

\appendix

\section{Implementation details} \label{section:appendix}

The computationally intensive components of the simulation are implemented in three C++ programs,
named \fitcat, \makegrid, and \photoz, while other tasks are primarily performed in interpreted
languages (Python, Perl, and IDL). The C++ programs are all parallelized with OpenMP and are
optimized for systems with non-uniform memory access (i.e., shared memory systems with multiple CPU
sockets or multiple memory controllers). These programs operate on input files that can trivially
be
split into smaller pieces so that the work can easily be distributed over the nodes of a computing
cluster. The IDL scripts are compatible with the free GNU Data Language interpreter, so no special
licenses need to be purchased in order to run the code. The simulation is customized to run
efficiently in Amazon's Elastic Compute Cloud, as described in \S\ref{section:pipeline-aws},
however it can also run on a single Linux workstation without modification.

Most of the adjustable parameters in the pipeline are stored in a single parameter file. To run new
simulations, the user typically only needs to modify the parameter file and possibly modify the
data
in the input files. The simulation code consists of 7 major steps:

\subsection*{Step 1: Input catalog model-fitting with \fitcat}

We begin by using \fitcat\ to assign SED model parameters (tuples of template type, $E(B-V)$,
and reddening law) to the objects in the COSMOS catalog, as discussed in
\S\ref{section:pipeline-input-fluxes}. For each unique redshift in the COSMOS catalog, a grid of
SED
models is created. The models are integrated against the COSMOS filters so that they can be
directly
compared with the fluxes in the catalog. In order to identify the model which best fits an object
in
the catalog, we begin by noting the redshift of the object. We then fetch the grid of models
corresponding to that redshift. The fluxes of each model in the grid are compared with the object's
fluxes and then the best-fitting model (i.e., the model which minimizes the $\chi^2$ difference) is
identified. The parameters of the best-fitting model are then stored in an output catalog, along
with the scaling factor by which the model was multiplied in order to minimize $\chi^2$. This step
in the simulation only needs to be performed when the template library is updated or a new catalog
becomes available.

\subsection*{Step 2: Instrument simulation}

In the second step, engineering data and survey properties, such as the integration times and
the physical dimensions of the instrument, are used to generate simulated LVF transmission curves
and PSF size. Noise characteristics for each band are also computed during this step and lead to
the full sky performances, described in \citet{Dore:2014}.

\subsection*{Step 3: Input photometry generation with \makegrid}

The \makegrid\ program computes fluxes by integrating filter profiles against templates after first
scaling, redshifting, and reddening the templates, as specified by an input file. Given 1) a
template library, 2) a set of \spherex\ LVF filter transmission curves, and 3) the list of model
parameters, identified in Step~1, \makegrid\ computes SEDs for the objects in the COSMOS catalog,
interpolated to the \spherex\ bands, as described in \S\ref{section:pipeline-input-fluxes}.

\subsection*{Step 4: Model grid generation with \makegrid}

We use \makegrid\ to generate a large grid of models of varying redshift, template type,
extinction, and reddening law, as described in \S\ref{section:photoz-algo}. The models are
integrated against the \spherex\ LVF transmission curves so that they can be directly compared with
the observed SEDs later (in Step~7). The model grid is stored in a compact binary file format,
optimized for fast reading.

This step is very computationally expensive. For example, a \spherex\
model grid containing $5.5\times10^7$ models requires 150 CPU core-hours, on an Intel Ivy Bridge
CPU, running at 2.7~GHz. Thus, this step is typically performed on a computing cluster while
another computer performs Steps~5 and 6. Note that model grids can be saved and stored for
later use; the grid only needs to be re-generated when changes are made to the instrument
specifications or the template library.

\subsection*{Step 5: Image generation}

For each of the \spherex\ LVF bands, a FITS image is generated, as described in
\S\ref{section:pipeline-images}, using the coordinate data in the COSMOS catalog, the fluxes
computed in Step~3, and the PSF determined in Step~2. The images are generated using utilities
included in IMCAT\footnote{\url{https://www.ifa.hawaii.edu/~kaiser/imcat/}}. The IMCAT utilities
are
not multi-threaded, so multiple processes are spawned simultaneously in order to accelerate the
image generation process.

\subsection*{Step 6: Observed SED generation}

The user can specify detection threshold magnitudes for various bands, based on the detection
limits
of the external survey that they are using to identify objects of interest. In the case of the
\spherex\ simulation, the detection thresholds are given by the depths of the \panstarrs\ PS1 and
WISE W1 and W2 bands. Objects that are brighter than at least one of these thresholds are marked as
detectable. An IDL script then performs photometry measurements on each FITS image, as described in
\S\ref{section:pipeline-photometry}, using the PSFs computed in Step~2 and the known coordinates of
the detectable objects. Once the fluxes have been measured, the data are re-organized to generate
observed SEDs for each object that was marked as detectable.

\subsection*{Step 7: Photometric redshift estimation with \photoz}

The \photoz\ program performs the photometric redshift estimation, described in
\S\ref{section:photoz-algo}. It reads the extracted SEDs (produced in Step~6) and the model grid,
(generated in Step~4). The program then produces a catalog containing estimated redshifts as well
as
additional information about each redshift estimate. For example, the skewness and kurtosis of the
liklihood distribution are reported along with the global minimum $\chi^2$ value and the ID number
of the associated best-fitting template.

The photometric redshift estimation algorithm is computationally expensive, particularly when large
model grids are used. For example, estimating the redshifts of $5.3\times10^4$ objects with SEDs
consisting of 100 bands, using a grid containing $5.5\times 10^7$ models requires approximately 260
CPU core-hours on a an Intel Ivy Bridge CPU, running at 2.7~GHz. The current implementation of
\photoz\ requires the entire model grid to be loaded into system memory on each individual
computing
node. Thus, the size of the model grid is currently limited by the amount of memory (RAM) available
in each individual node of the cluster. In a future (production-quality) version of the redshift
estimation pipeline, the grid will likely be broken into several redshift-based partitions that are
small enough to fit into the memory of a graphics processing unit (GPU). A cluster of GPUs would
then compute the elements of the redshift liklihood distributions; the moments of the distributions
would still be computed using a CPU-native code.

\bibliographystyle{yahapj} 
\bibliography{refs}

\begin{thebibliography}{}
\providecommand\natexlab[1]{#1}
\providecommand\JournalTitle[1]{#1}

\bibitem[{{Allen}(1976)}]{allen1976}
{Allen}, D.~A. 1976,
  \href{http://dx.doi.org/10.1093/mnras/174.1.29P}{\JournalTitle{\mnras}, 174,
  29P}

\bibitem[{{Arnouts} {et~al.}(1999){Arnouts}, {Cristiani}, {Moscardini},
  {Matarrese}, {Lucchin}, {Fontana}, \& {Giallongo}}]{lephare1}
{Arnouts}, S., {Cristiani}, S., {Moscardini}, L., {et~al.} 1999,
  \href{http://dx.doi.org/10.1046/j.1365-8711.1999.02978.x}{\JournalTitle{\mnras},
  310, 540}

\bibitem[{{Arnouts} \& {Ilbert}(2011)}]{lephare2}
{Arnouts}, S., \& {Ilbert}, O. 2011, {LePHARE: Photometric Analysis for
  Redshift Estimate}, Astrophysics Source Code Library,
  \href{http://arxiv.org/abs/1108.009}{{\sffamily ascl:1108.009}}

\bibitem[{{Brown} {et~al.}(2014){Brown}, {Moustakas}, {Smith}, {da Cunha},
  {Jarrett}, {Imanishi}, {Armus}, {Brandl}, \& {Peek}}]{browntemplates}
{Brown}, M.~J.~I., {Moustakas}, J., {Smith}, J.-D.~T., {et~al.} 2014,
  \href{http://dx.doi.org/10.1088/0067-0049/212/2/18}{\JournalTitle{\apjs},
  212, 18}

\bibitem[{{Bruzual} \& {Charlot}(2003)}]{bruzual-charlot}
{Bruzual}, G., \& {Charlot}, S. 2003,
  \href{http://dx.doi.org/10.1046/j.1365-8711.2003.06897.x}{\JournalTitle{\mnras},
  344, 1000}

\bibitem[{{Calzetti} {et~al.}(2000){Calzetti}, {Armus}, {Bohlin}, {Kinney},
  {Koornneef}, \& {Storchi-Bergmann}}]{calzetti2000}
{Calzetti}, D., {Armus}, L., {Bohlin}, R.~C., {et~al.} 2000,
  \href{http://dx.doi.org/10.1086/308692}{\JournalTitle{\apj}, 533, 682}

\bibitem[{{Capak} {et~al.}(2007){Capak}, {Aussel}, {Ajiki}, {McCracken},
  {Mobasher}, {Scoville}, {Shopbell}, {Taniguchi}, {Thompson}, {Tribiano},
  {Sasaki}, {Blain}, {Brusa}, {Carilli}, {Comastri}, {Carollo}, {Cassata},
  {Colbert}, {Ellis}, {Elvis}, {Giavalisco}, {Green}, {Guzzo}, {Hasinger},
  {Ilbert}, {Impey}, {Jahnke}, {Kartaltepe}, {Kneib}, {Koda}, {Koekemoer},
  {Komiyama}, {Leauthaud}, {Le Fevre}, {Lilly}, {Liu}, {Massey}, {Miyazaki},
  {Murayama}, {Nagao}, {Peacock}, {Pickles}, {Porciani}, {Renzini}, {Rhodes},
  {Rich}, {Salvato}, {Sanders}, {Scarlata}, {Schiminovich}, {Schinnerer},
  {Scodeggio}, {Sheth}, {Shioya}, {Tasca}, {Taylor}, {Yan}, \&
  {Zamorani}}]{capak2007}
{Capak}, P., {Aussel}, H., {Ajiki}, M., {et~al.} 2007,
  \href{http://dx.doi.org/10.1086/519081}{\JournalTitle{\apjs}, 172, 99}

\bibitem[{{Chabrier} {et~al.}(2000){Chabrier}, {Baraffe}, {Allard}, \&
  {Hauschildt}}]{Chabrier:2000}
{Chabrier}, G., {Baraffe}, I., {Allard}, F., \& {Hauschildt}, P. 2000,
  \href{http://dx.doi.org/10.1086/309513}{\JournalTitle{\apj}, 542, 464}

\bibitem[{{Coil} {et~al.}(2011){Coil}, {Blanton}, {Burles}, {Cool},
  {Eisenstein}, {Moustakas}, {Wong}, {Zhu}, {Aird}, {Bernstein}, {Bolton}, \&
  {Hogg}}]{primus1}
{Coil}, A.~L., {Blanton}, M.~R., {Burles}, S.~M., {et~al.} 2011,
  \href{http://dx.doi.org/10.1088/0004-637X/741/1/8}{\JournalTitle{\apj}, 741,
  8}

\bibitem[{{Cool} {et~al.}(2013){Cool}, {Moustakas}, {Blanton}, {Burles},
  {Coil}, {Eisenstein}, {Wong}, {Zhu}, {Aird}, {Bernstein}, {Bolton}, {Hogg},
  \& {Mendez}}]{primus2}
{Cool}, R.~J., {Moustakas}, J., {Blanton}, M.~R., {et~al.} 2013,
  \href{http://dx.doi.org/10.1088/0004-637X/767/2/118}{\JournalTitle{\apj},
  767, 118}

\bibitem[{{da Cunha} {et~al.}(2012){da Cunha}, {Charlot}, {Dunne}, {Smith}, \&
  {Rowlands}}]{magphys}
{da Cunha}, E., {Charlot}, S., {Dunne}, L., {Smith}, D., \& {Rowlands}, K.
  2012, \href{http://dx.doi.org/10.1017/S1743921312009283}{in IAU Symposium,
  Vol. 284, The Spectral Energy Distribution of Galaxies - SED 2011, ed. R.~J.
  {Tuffs} \& C.~C. {Popescu}}, 292

\bibitem[{{Dor{\'e}} {et~al.}(2014){Dor{\'e}}, {Bock}, {Ashby}, {Capak},
  {Cooray}, {de Putter}, {Eifler}, {Flagey}, {Gong}, {Habib}, {Heitmann},
  {Hirata}, {Jeong}, {Katti}, {Korngut}, {Krause}, {Lee}, {Masters},
  {Mauskopf}, {Melnick}, {Mennesson}, {Nguyen}, {{\"O}berg}, {Pullen},
  {Raccanelli}, {Smith}, {Song}, {Tolls}, {Unwin}, {Venumadhav}, {Viero},
  {Werner}, \& {Zemcov}}]{Dore:2014}
{Dor{\'e}}, O., {Bock}, J., {Ashby}, M., {et~al.} 2014, \JournalTitle{ArXiv
  e-prints}, \href{http://arxiv.org/abs/1412.4872}{{\sffamily arXiv:1412.4872}}

\bibitem[{{Elliott} {et~al.}(2016){Elliott}, {McComas}, {Valek}, {Nicolaou},
  {Weidner}, \& {Livadiotis}}]{newhorizons}
{Elliott}, H.~A., {McComas}, D.~J., {Valek}, P., {et~al.} 2016,
  \href{http://dx.doi.org/10.3847/0067-0049/223/2/19}{\JournalTitle{\apjs},
  223, 19}

\bibitem[{{Fitzpatrick} \& {Massa}(1986)}]{fitzpatrick1986}
{Fitzpatrick}, E.~L., \& {Massa}, D. 1986,
  \href{http://dx.doi.org/10.1086/164415}{\JournalTitle{\apj}, 307, 286}

\bibitem[{{Ilbert} {et~al.}(2009){Ilbert}, {Capak}, {Salvato}, {Aussel},
  {McCracken}, {Sanders}, {Scoville}, {Kartaltepe}, {Arnouts}, {Le Floc'h},
  {Mobasher}, {Taniguchi}, {Lamareille}, {Leauthaud}, {Sasaki}, {Thompson},
  {Zamojski}, {Zamorani}, {Bardelli}, {Bolzonella}, {Bongiorno}, {Brusa},
  {Caputi}, {Carollo}, {Contini}, {Cook}, {Coppa}, {Cucciati}, {de la Torre},
  {de Ravel}, {Franzetti}, {Garilli}, {Hasinger}, {Iovino}, {Kampczyk},
  {Kneib}, {Knobel}, {Kovac}, {Le Borgne}, {Le Brun}, {F{\`e}vre}, {Lilly},
  {Looper}, {Maier}, {Mainieri}, {Mellier}, {Mignoli}, {Murayama}, {Pell{\`o}},
  {Peng}, {P{\'e}rez-Montero}, {Renzini}, {Ricciardelli}, {Schiminovich},
  {Scodeggio}, {Shioya}, {Silverman}, {Surace}, {Tanaka}, {Tasca}, {Tresse},
  {Vergani}, \& {Zucca}}]{ilbert2009}
{Ilbert}, O., {Capak}, P., {Salvato}, M., {et~al.} 2009,
  \href{http://dx.doi.org/10.1088/0004-637X/690/2/1236}{\JournalTitle{\apj},
  690, 1236}

\bibitem[{{Kaiser}(2004)}]{panstarrs2004}
{Kaiser}, N. 2004, \href{http://dx.doi.org/10.1117/12.552472}{in \procspie,
  Vol. 5489, Ground-based Telescopes, ed. J.~M. {Oschmann}, Jr.}, 11

\bibitem[{{Kaiser} {et~al.}(2002){Kaiser}, {Aussel}, {Burke}, {Boesgaard},
  {Chambers}, {Chun}, {Heasley}, {Hodapp}, {Hunt}, {Jedicke}, {Jewitt},
  {Kudritzki}, {Luppino}, {Maberry}, {Magnier}, {Monet}, {Onaka}, {Pickles},
  {Rhoads}, {Simon}, {Szalay}, {Szapudi}, {Tholen}, {Tonry}, {Waterson}, \&
  {Wick}}]{panstarrs2002}
{Kaiser}, N., {Aussel}, H., {Burke}, B.~E., {et~al.} 2002,
  \href{http://dx.doi.org/10.1117/12.457365}{in \procspie, Vol. 4836, Survey
  and Other Telescope Technologies and Discoveries, ed. J.~A. {Tyson} \&
  S.~{Wolff}}, 154

\bibitem[{{Koekemoer} {et~al.}(2011){Koekemoer}, {Faber}, {Ferguson}, {Grogin},
  {Kocevski}, {Koo}, {Lai}, {Lotz}, {Lucas}, {McGrath}, {Ogaz}, {Rajan},
  {Riess}, {Rodney}, {Strolger}, {Casertano}, {Castellano}, {Dahlen},
  {Dickinson}, {Dolch}, {Fontana}, {Giavalisco}, {Grazian}, {Guo}, {Hathi},
  {Huang}, {van der Wel}, {Yan}, {Acquaviva}, {Alexander}, {Almaini}, {Ashby},
  {Barden}, {Bell}, {Bournaud}, {Brown}, {Caputi}, {Cassata}, {Challis},
  {Chary}, {Cheung}, {Cirasuolo}, {Conselice}, {Roshan Cooray}, {Croton},
  {Daddi}, {Dav{\'e}}, {de Mello}, {de Ravel}, {Dekel}, {Donley}, {Dunlop},
  {Dutton}, {Elbaz}, {Fazio}, {Filippenko}, {Finkelstein}, {Frazer}, {Gardner},
  {Garnavich}, {Gawiser}, {Gruetzbauch}, {Hartley}, {H{\"a}ussler},
  {Herrington}, {Hopkins}, {Huang}, {Jha}, {Johnson}, {Kartaltepe},
  {Khostovan}, {Kirshner}, {Lani}, {Lee}, {Li}, {Madau}, {McCarthy},
  {McIntosh}, {McLure}, {McPartland}, {Mobasher}, {Moreira}, {Mortlock},
  {Moustakas}, {Mozena}, {Nandra}, {Newman}, {Nielsen}, {Niemi}, {Noeske},
  {Papovich}, {Pentericci}, {Pope}, {Primack}, {Ravindranath}, {Reddy},
  {Renzini}, {Rix}, {Robaina}, {Rosario}, {Rosati}, {Salimbeni}, {Scarlata},
  {Siana}, {Simard}, {Smidt}, {Snyder}, {Somerville}, {Spinrad}, {Straughn},
  {Telford}, {Teplitz}, {Trump}, {Vargas}, {Villforth}, {Wagner}, {Wandro},
  {Wechsler}, {Weiner}, {Wiklind}, {Wild}, {Wilson}, {Wuyts}, \&
  {Yun}}]{candels}
{Koekemoer}, A.~M., {Faber}, S.~M., {Ferguson}, H.~C., {et~al.} 2011,
  \href{http://dx.doi.org/10.1088/0067-0049/197/2/36}{\JournalTitle{\apjs},
  197, 36}

\bibitem[{{Laigle} {et~al.}(2016){Laigle}, {McCracken}, {Ilbert}, {Hsieh},
  {Davidzon}, {Capak}, {Hasinger}, {Silverman}, {Pichon}, {Coupon}, {Aussel},
  {Le Borgne}, {Caputi}, {Cassata}, {Chang}, {Civano}, {Dunlop}, {Fynbo},
  {Kartaltepe}, {Koekemoer}, {Le F{\`e}vre}, {Le Floc'h}, {Leauthaud}, {Lilly},
  {Lin}, {Marchesi}, {Milvang-Jensen}, {Salvato}, {Sanders}, {Scoville},
  {Smolcic}, {Stockmann}, {Taniguchi}, {Tasca}, {Toft}, {Vaccari}, \&
  {Zabl}}]{laigle2016}
{Laigle}, C., {McCracken}, H.~J., {Ilbert}, O., {et~al.} 2016,
  \href{http://dx.doi.org/10.3847/0067-0049/224/2/24}{\JournalTitle{\apjs},
  224, 24}

\bibitem[{{Laureijs} {et~al.}(2011){Laureijs}, {Amiaux}, {Arduini},
  {Augu{\`e}res}, {Brinchmann}, {Cole}, {Cropper}, {Dabin}, {Duvet}, {Ealet},
  \& et~al.}]{Laureijs:2011gra}
{Laureijs}, R., {Amiaux}, J., {Arduini}, S., {et~al.} 2011, \JournalTitle{ArXiv
  e-prints}, \href{http://arxiv.org/abs/1110.3193}{{\sffamily arXiv:1110.3193
  [astro-ph.CO]}}

\bibitem[{{Levi} {et~al.}(2013){Levi}, {Bebek}, {Beers}, {Blum}, {Cahn},
  {Eisenstein}, {Flaugher}, {Honscheid}, {Kron}, {Lahav}, {McDonald}, {Roe},
  {Schlegel}, \& {representing the DESI collaboration}}]{Levi:2013}
{Levi}, M., {Bebek}, C., {Beers}, T., {et~al.} 2013, \JournalTitle{ArXiv
  e-prints}, \href{http://arxiv.org/abs/1308.0847}{{\sffamily arXiv:1308.0847
  [astro-ph.CO]}}

\bibitem[{{LSST Science Collaboration} {et~al.}(2009){LSST Science
  Collaboration}, {Abell}, {Allison}, {Anderson}, {Andrew}, {Angel}, {Armus},
  {Arnett}, {Asztalos}, {Axelrod}, \& et~al.}]{SciBook}
{LSST Science Collaboration}, {Abell}, P.~A., {Allison}, J., {et~al.} 2009,
  \JournalTitle{ArXiv e-prints},
  \href{http://arxiv.org/abs/0912.0201}{{\sffamily arXiv:0912.0201
  [astro-ph.IM]}}

\bibitem[{{Masters} {et~al.}(2015){Masters}, {Capak}, {Stern}, {Ilbert},
  {Salvato}, {Schmidt}, {Longo}, {Rhodes}, {Paltani}, {Mobasher}, {Hoekstra},
  {Hildebrandt}, {Coupon}, {Steinhardt}, {Speagle}, {Faisst}, {Kalinich},
  {Brodwin}, {Brescia}, \& {Cavuoti}}]{masters2015}
{Masters}, D., {Capak}, P., {Stern}, D., {et~al.} 2015,
  \href{http://dx.doi.org/10.1088/0004-637X/813/1/53}{\JournalTitle{\apj}, 813,
  53}

\bibitem[{{McCracken} {et~al.}(2013){McCracken}, {Milvang-Jensen}, {Dunlop},
  {Franx}, {Fynbo}, {Le F{\`e}vre}, {Holt}, {Caputi}, {Goranova}, {Buitrago},
  {Emerson}, {Freudling}, {Herent}, {Hudelot}, {L{\'o}pez-Sanjuan}, {Magnard},
  {Muzzin}, {Mellier}, {M{\o}ller}, {Nilsson}, {Sutherland}, {Tasca}, \&
  {Zabl}}]{mccracken2013}
{McCracken}, H.~J., {Milvang-Jensen}, B., {Dunlop}, J., {et~al.} 2013,
  \JournalTitle{The Messenger}, 154, 29

\bibitem[{{Moles} {et~al.}(2008){Moles}, {Ben{\'{\i}}tez}, {Aguerri}, {Alfaro},
  {Broadhurst}, {Cabrera-Ca{\~n}o}, {Castander}, {Cepa}, {Cervi{\~n}o},
  {Crist{\'o}bal-Hornillos}, {Fern{\'a}ndez-Soto}, {Gonz{\'a}lez Delgado},
  {Infante}, {M{\'a}rquez}, {Mart{\'{\i}}nez}, {Masegosa}, {del Olmo}, {Perea},
  {Prada}, {Quintana}, \& {S{\'a}nchez}}]{alhambra}
{Moles}, M., {Ben{\'{\i}}tez}, N., {Aguerri}, J.~A.~L., {et~al.} 2008,
  \href{http://dx.doi.org/10.1088/0004-6256/136/3/1325}{\JournalTitle{\aj},
  136, 1325}

\bibitem[{{Neugebauer} {et~al.}(1984){Neugebauer}, {Habing}, {van Duinen},
  {Aumann}, {Baud}, {Beichman}, {Beintema}, {Boggess}, {Clegg}, {de Jong},
  {Emerson}, {Gautier}, {Gillett}, {Harris}, {Hauser}, {Houck}, {Jennings},
  {Low}, {Marsden}, {Miley}, {Olnon}, {Pottasch}, {Raimond}, {Rowan-Robinson},
  {Soifer}, {Walker}, {Wesselius}, \& {Young}}]{iras}
{Neugebauer}, G., {Habing}, H.~J., {van Duinen}, R., {et~al.} 1984,
  \href{http://dx.doi.org/10.1086/184209}{\JournalTitle{\apjl}, 278, L1}

\bibitem[{{Prevot} {et~al.}(1984){Prevot}, {Lequeux}, {Prevot}, {Maurice}, \&
  {Rocca-Volmerange}}]{prevot1984}
{Prevot}, M.~L., {Lequeux}, J., {Prevot}, L., {Maurice}, E., \&
  {Rocca-Volmerange}, B. 1984, \JournalTitle{\aap}, 132, 389

\bibitem[{{Salvato} {et~al.}(2011){Salvato}, {Ilbert}, {Hasinger}, {Rau},
  {Civano}, {Zamorani}, {Brusa}, {Elvis}, {Vignali}, {Aussel}, {Comastri},
  {Fiore}, {Le Floc'h}, {Mainieri}, {Bardelli}, {Bolzonella}, {Bongiorno},
  {Capak}, {Caputi}, {Cappelluti}, {Carollo}, {Contini}, {Garilli}, {Iovino},
  {Fotopoulou}, {Fruscione}, {Gilli}, {Halliday}, {Kneib}, {Kakazu},
  {Kartaltepe}, {Koekemoer}, {Kovac}, {Ideue}, {Ikeda}, {Impey}, {Le Fevre},
  {Lamareille}, {Lanzuisi}, {Le Borgne}, {Le Brun}, {Lilly}, {Maier},
  {Manohar}, {Masters}, {McCracken}, {Messias}, {Mignoli}, {Mobasher}, {Nagao},
  {Pello}, {Puccetti}, {Perez-Montero}, {Renzini}, {Sargent}, {Sanders},
  {Scodeggio}, {Scoville}, {Shopbell}, {Silvermann}, {Taniguchi}, {Tasca},
  {Tresse}, {Trump}, \& {Zucca}}]{salvato2011}
{Salvato}, M., {Ilbert}, O., {Hasinger}, G., {et~al.} 2011,
  \href{http://dx.doi.org/10.1088/0004-637X/742/2/61}{\JournalTitle{\apj}, 742,
  61}

\bibitem[{{Scoville} {et~al.}(2007){Scoville}, {Aussel}, {Brusa}, {Capak},
  {Carollo}, {Elvis}, {Giavalisco}, {Guzzo}, {Hasinger}, {Impey}, {Kneib},
  {LeFevre}, {Lilly}, {Mobasher}, {Renzini}, {Rich}, {Sanders}, {Schinnerer},
  {Schminovich}, {Shopbell}, {Taniguchi}, \& {Tyson}}]{scoville2007}
{Scoville}, N., {Aussel}, H., {Brusa}, M., {et~al.} 2007,
  \href{http://dx.doi.org/10.1086/516585}{\JournalTitle{\apjs}, 172, 1}

\bibitem[{{Seaton}(1979)}]{seaton1979}
{Seaton}, M.~J. 1979,
  \href{http://dx.doi.org/10.1093/mnras/187.1.73P}{\JournalTitle{\mnras}, 187,
  73P}

\bibitem[{{Smoot} {et~al.}(1992){Smoot}, {Bennett}, {Kogut}, {Wright}, {Aymon},
  {Boggess}, {Cheng}, {de Amici}, {Gulkis}, {Hauser}, {Hinshaw}, {Jackson},
  {Janssen}, {Kaita}, {Kelsall}, {Keegstra}, {Lineweaver}, {Loewenstein},
  {Lubin}, {Mather}, {Meyer}, {Moseley}, {Murdock}, {Rokke}, {Silverberg},
  {Tenorio}, {Weiss}, \& {Wilkinson}}]{cobe}
{Smoot}, G.~F., {Bennett}, C.~L., {Kogut}, A., {et~al.} 1992,
  \href{http://dx.doi.org/10.1086/186504}{\JournalTitle{\apjl}, 396, L1}

\bibitem[{{Spergel} {et~al.}(2015){Spergel}, {Gehrels}, {Baltay}, {Bennett},
  {Breckinridge}, {Donahue}, {Dressler}, {Gaudi}, {Greene}, {Guyon}, {Hirata},
  {Kalirai}, {Kasdin}, {Macintosh}, {Moos}, {Perlmutter}, {Postman},
  {Rauscher}, {Rhodes}, {Wang}, {Weinberg}, {Benford}, {Hudson}, {Jeong},
  {Mellier}, {Traub}, {Yamada}, {Capak}, {Colbert}, {Masters}, {Penny},
  {Savransky}, {Stern}, {Zimmerman}, {Barry}, {Bartusek}, {Carpenter}, {Cheng},
  {Content}, {Dekens}, {Demers}, {Grady}, {Jackson}, {Kuan}, {Kruk}, {Melton},
  {Nemati}, {Parvin}, {Poberezhskiy}, {Peddie}, {Ruffa}, {Wallace}, {Whipple},
  {Wollack}, \& {Zhao}}]{Spergel:2015sza}
{Spergel}, D., {Gehrels}, N., {Baltay}, C., {et~al.} 2015, \JournalTitle{ArXiv
  e-prints}, \href{http://arxiv.org/abs/1503.03757}{{\sffamily arXiv:1503.03757
  [astro-ph.IM]}}

\bibitem[{{Stickley} \& {Aragon-Calvo}(2015)}]{stratos}
{Stickley}, N.~R., \& {Aragon-Calvo}, M.~A. 2015, \JournalTitle{ArXiv
  e-prints}, \href{http://arxiv.org/abs/1503.02233}{{\sffamily arXiv:1503.02233
  [astro-ph.IM]}}

\bibitem[{{Takada} {et~al.}(2014){Takada}, {Ellis}, {Chiba}, {Greene},
  {Aihara}, {Arimoto}, {Bundy}, {Cohen}, {Dor{\'e}}, {Graves}, {Gunn},
  {Heckman}, {Hirata}, {Ho}, {Kneib}, {F{\`e}vre}, {Lin}, {More}, {Murayama},
  {Nagao}, {Ouchi}, {Seiffert}, {Silverman}, {Sodr{\'e}}, {Spergel}, {Strauss},
  {Sugai}, {Suto}, {Takami}, \& {Wyse}}]{Takada:2014}
{Takada}, M., {Ellis}, R.~S., {Chiba}, M., {et~al.} 2014,
  \href{http://dx.doi.org/10.1093/pasj/pst019}{\JournalTitle{\pasj}, 66, R1}

\bibitem[{{Wright} {et~al.}(2010){Wright}, {Eisenhardt}, {Mainzer}, {Ressler},
  {Cutri}, {Jarrett}, {Kirkpatrick}, {Padgett}, {McMillan}, {Skrutskie},
  {Stanford}, {Cohen}, {Walker}, {Mather}, {Leisawitz}, {Gautier}, {McLean},
  {Benford}, {Lonsdale}, {Blain}, {Mendez}, {Irace}, {Duval}, {Liu}, {Royer},
  {Heinrichsen}, {Howard}, {Shannon}, {Kendall}, {Walsh}, {Larsen}, {Cardon},
  {Schick}, {Schwalm}, {Abid}, {Fabinsky}, {Naes}, \& {Tsai}}]{wise}
{Wright}, E.~L., {Eisenhardt}, P.~R.~M., {Mainzer}, A.~K., {et~al.} 2010,
  \href{http://dx.doi.org/10.1088/0004-6256/140/6/1868}{\JournalTitle{\aj},
  140, 1868}

\end{thebibliography}

\end{document}